\documentclass[prb,twocolumn,preprintnumbers,amsmath,amssymb,groupaddress]{revtex4}

\usepackage{graphicx}
\usepackage{bm}

\begin{document}

\title{Dynamical density-density correlations in the one-dimensional Bose gas}

\author{Jean-S\'ebastien Caux} 
\author{Pasquale Calabrese} 
\affiliation{Institute for Theoretical Physics, University of
 Amsterdam, 
1018 XE Amsterdam, The Netherlands.}

\date{\today}

\begin{abstract}
The zero-temperature 
dynamical structure factor of the one-dimensional Bose gas with delta-function
interaction (Lieb-Liniger model) is computed as a function of momentum and
frequency 
using a hybrid theoretical/numerical method based on the exact 
Bethe Ansatz solution.  This 
allows to interpolate continuously between the weakly-coupled Thomas-Fermi
and strongly-coupled Tonks-Girardeau regimes.  The results should be
experimentally accessible
with Bragg spectroscopy.
\end{abstract}

\maketitle

The physics of low dimensional atomic systems presents very special features
as compared to the three-dimensional case.  As the temperature
is lowered, a uniform gas of bosons in three dimensions will undergo a 
transition to a Bose-Einstein condensate (BEC) \cite{PitaevskiiBOOK};  in the one-dimensional
case low-energy fluctuations prevent long-range order.  
For trapped gases, the situation changes and 
three regimes become possible in 1D\cite{PetrovPRL85}:
true condensate, quasicondensate, and a strongly-interacting regime, 
with BEC limited to extremely small interaction between particles.
Trapped 1D gases are now accessible experimentally \cite{GoerlitzPRL87,GreinerPRL87,MoritzPRL91}
in all regimes, the most challenging to obtain being the strongly-interacting case 
\cite{ParedesNATURE429,KinoshitaSCIENCE305}, which can survive 
without fast decay due to a reduced three-body recombination rate 
\cite{GangardtPRL90,LaburthePRL92} (a consequence of fermionization).  

A natural starting point for the theoretical description of one-dimensional atomic gases 
in this last regime is provided by bosons with delta-function interaction (the Lieb-Liniger model 
\cite{LiebPR130}), whose Hamiltonian is given by
\begin{equation}
H = -\sum_{j=1}^N \frac{\partial^2}{\partial_{x_j}^2} + 2c \sum_{\langle i,j \rangle} \delta(x_i - x_j)
\label{LL}
\end{equation}
in which $c > 0$ is the coupling constant, and the sum is over pairs
(we have put $\hbar = 1 = 2m$ for simplicity).  For definiteness, we consider
a system of length $L$ with periodic boundary conditions.
In the thermodynamic limit, 
the physics of the model depends
on a single parameter $\gamma = c/n$ where $n = N/L$ is the particle density.  
In 1D, in stark contrast to higher dimensions, low densities lead one to 
the strong-coupling regime of impenetrable bosons, 
know as the Tonks-Girardeau \cite{TonksPR50,GirardeauJMP1} limit.

Although equilibrium thermodynamic properties of the Lieb-Liniger model are accessible
via the Bethe Ansatz \cite{YangJMP10}, dynamical objects such as 
correlation functions cannot be readily obtained with this scheme.  
For example, the zero-temperature density-density correlation function
(written here in Fourier space, where it is also known as 
the dynamical structure factor (DSF)),
\begin{equation}
S (k, \omega) = \int_0^L dx \int dt ~e^{-i k x + i \omega t}
\langle \rho (x, t) \rho (0, 0) \rangle
\label{DSF}
\end{equation}
(in which $\rho (x) = \sum_{j=1}^N \delta(x - x_j)$) 
has up to now resisted all efforts towards an exact computation.  The
present paper presents a reliable and efficient method for computing this,
based on mixing integrability and numerics.

Many approximate theoretical schemes have been developed to tackle this
issue.  In the BEC regime, 
Bogoliubov theory can be used in conjunction with local density,
impulse or eikonal approximations \cite{ZambelliPRA61}.  
Specifically in 1D, 
an effective harmonic fluid
approach (Luttinger liquid theory) \cite{HaldanePRL47} can be used
to obtain information on the 
asymptotics of static and dynamical correlation functions
at zero and nonzero temperature \cite{BerkovichPLA142,CastroNetoPRB50,LuxatPRA67}.
Inversely to asymptotics, a small distance Taylor expansion was also proposed
\cite{OlshaniiPRL91}.  Yet another possibility is to exploit an exact fermion mapping, 
and use the Hartree-Fock and generalized random phase approximation to get
dynamical correlators near the Tonks-Girardeau limit in a $1/\gamma$ expansion \cite{BrandPRA72_73}.
Quantum Monte Carlo has been used to study this limit \cite{PolletPRL93}, and to numerically obtain the 
pair distribution function and static structure factor \cite{AstrakharchikPRA68}.
However, up to now, there is no overall reliable method for obtaining the full
momentum and frequency dependence of the DSF.

In view of the integrability of (\ref{LL}), one could expect to obtain nonperturbative
results for objects such as (\ref{DSF}).  Much recent
progress on the computation of correlation functions for this and other 1D quantum
integrable models has in fact been achieved through the Algebraic Bethe Ansatz 
\cite{KorepinBOOK,KorepinCMP94,SlavnovTMP79_82}.  In this paper, we wish to present a novel 
method for obtaining dynamical
correlation functions of model (\ref{LL}), which is based on these developments.
We will obtain the dynamical structure factor for finite but large systems, 
starting directly from the Bethe Ansatz solution, in a way which
is reminiscent of recent work by one of us on dynamical spin-spin correlation 
functions in Heisenberg
magnets \cite{CauxPRL95}.  In particular, the momentum and frequency dependence
of the DSF is fully characterized by our approach.
All our results are presented in Figures 1-3.
The static structure factor $S(k) = \int \frac{d\omega}{2\pi} S(k, \omega)$
is also obtained as a subset of our results.
The DSF itself is 
experimentally accessible through Fourier sampling of time-of-flight
images \cite{DuanPRL96} or through Bragg spectroscopy \cite{StengerPRL82}.

\begin{figure}[!]
\begin{tabular}{ll}
\includegraphics[width=4.3cm]{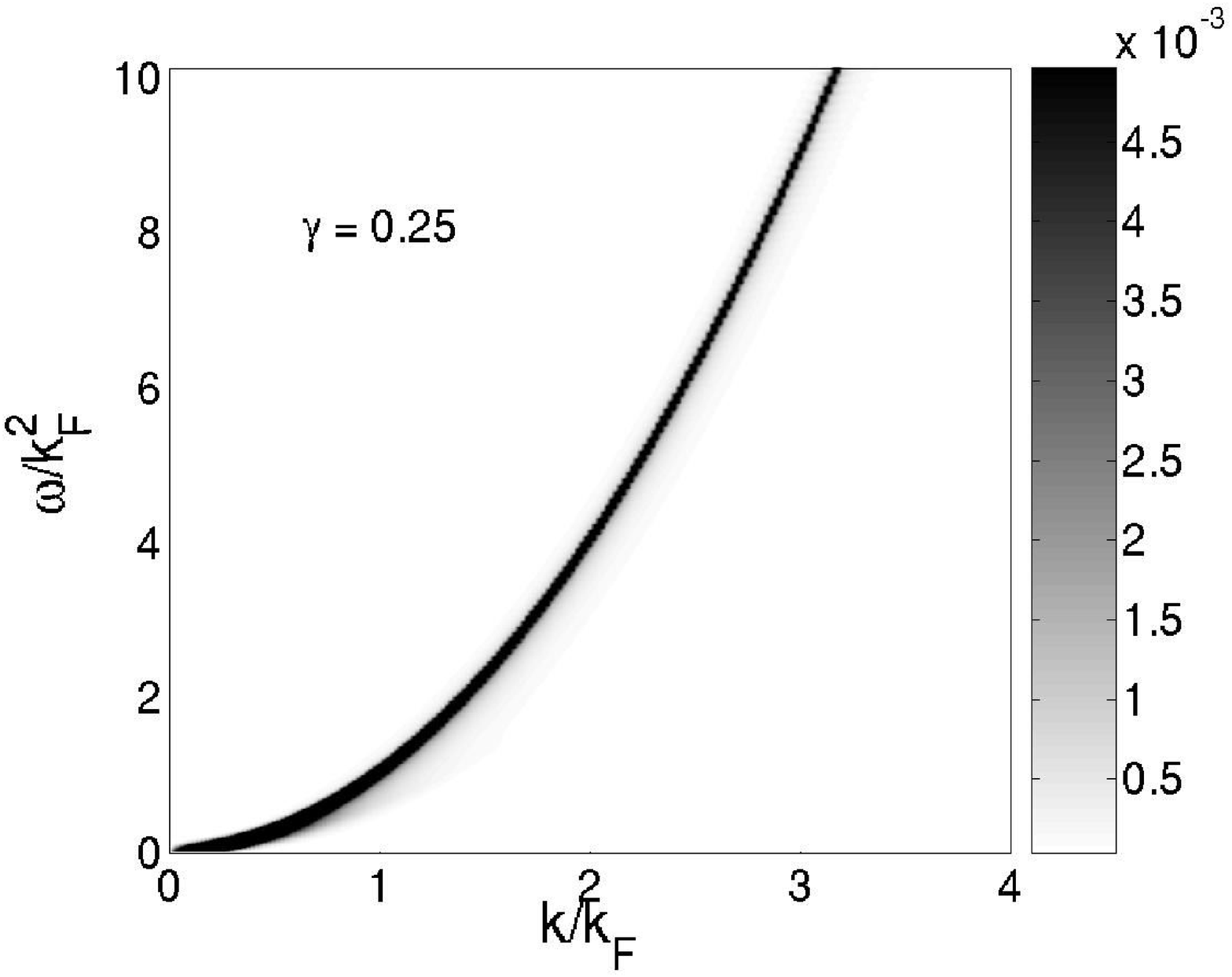}
&
\includegraphics[width=4.3cm]{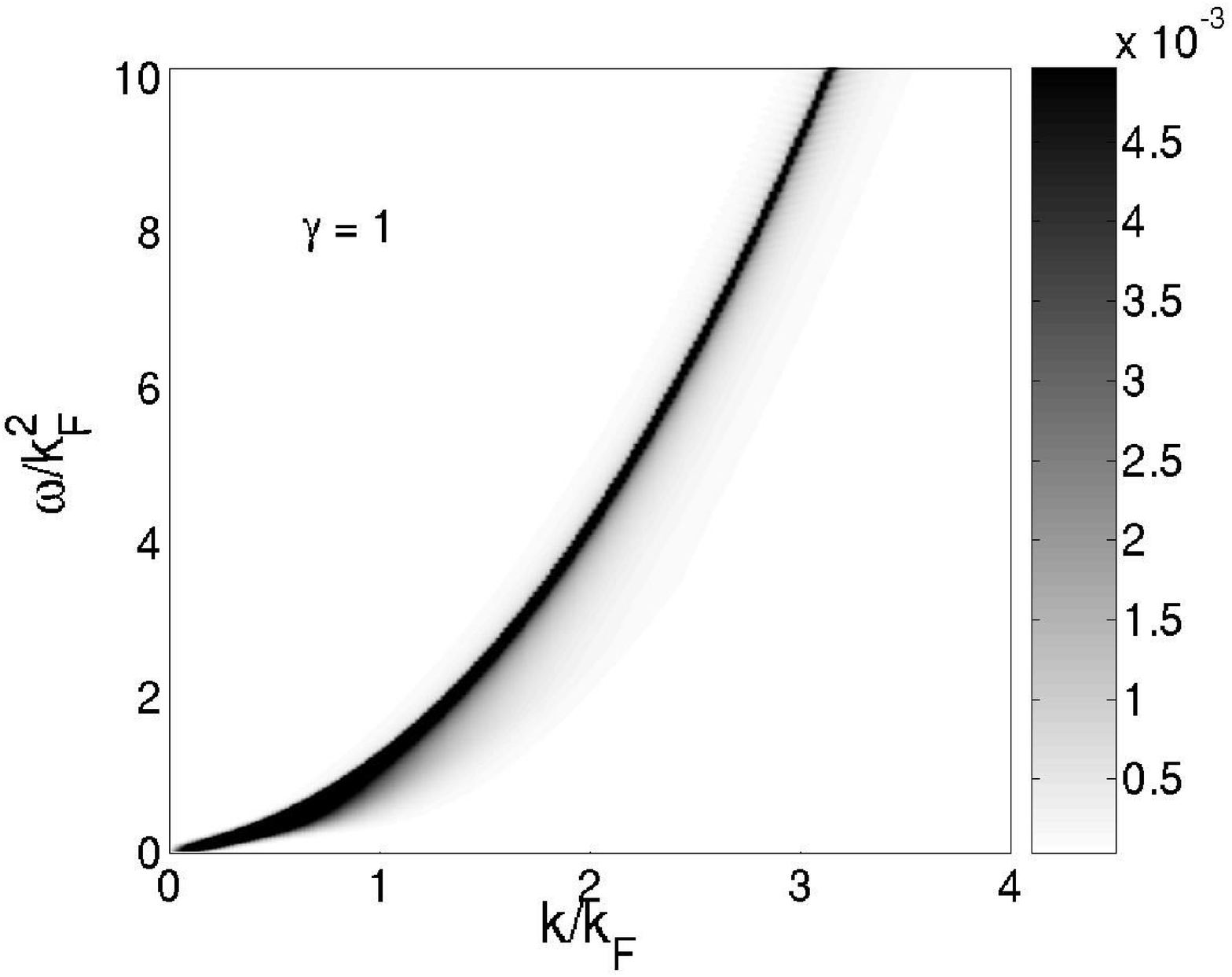} \\
\includegraphics[width=4.3cm]{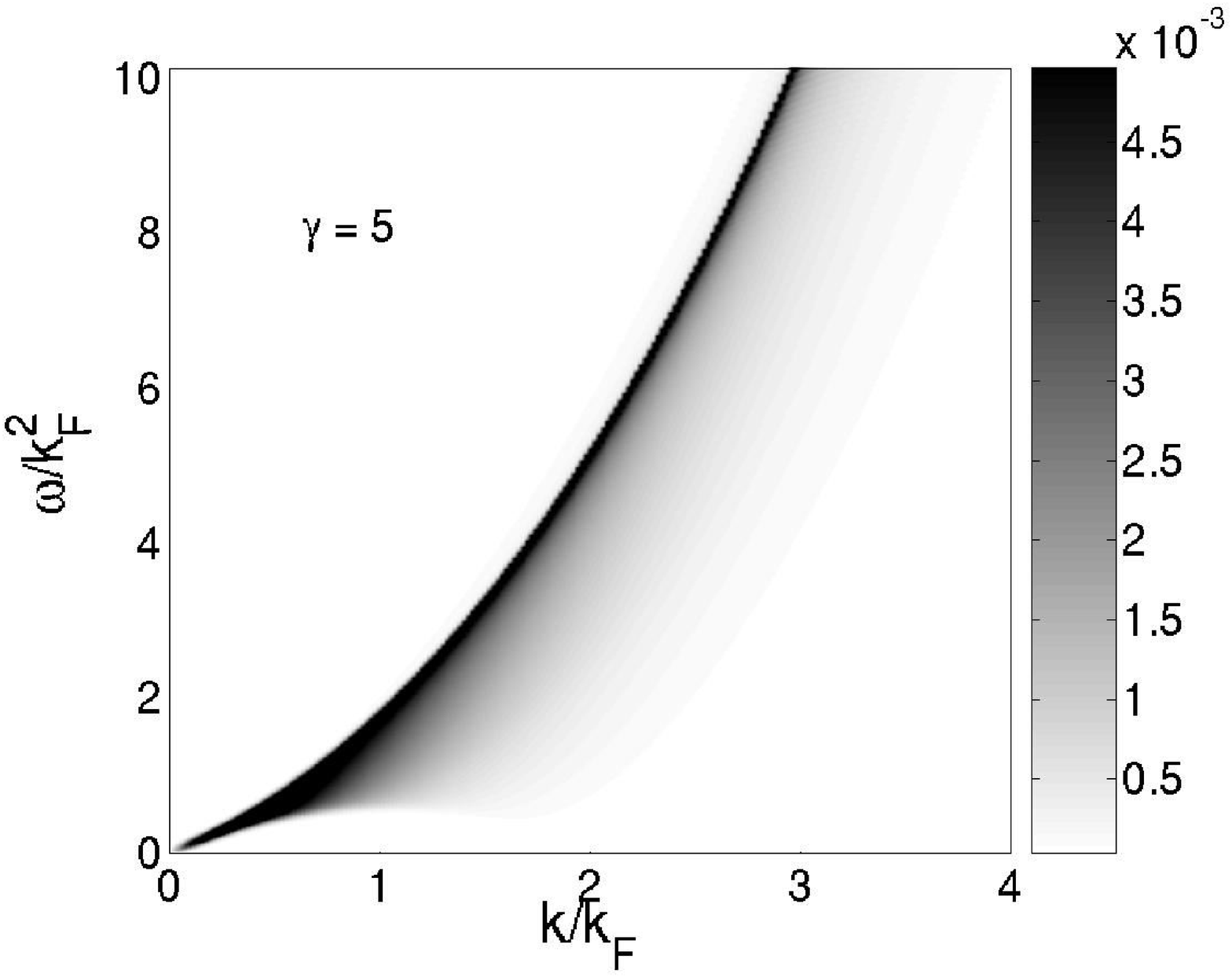}
&
\includegraphics[width=4.3cm]{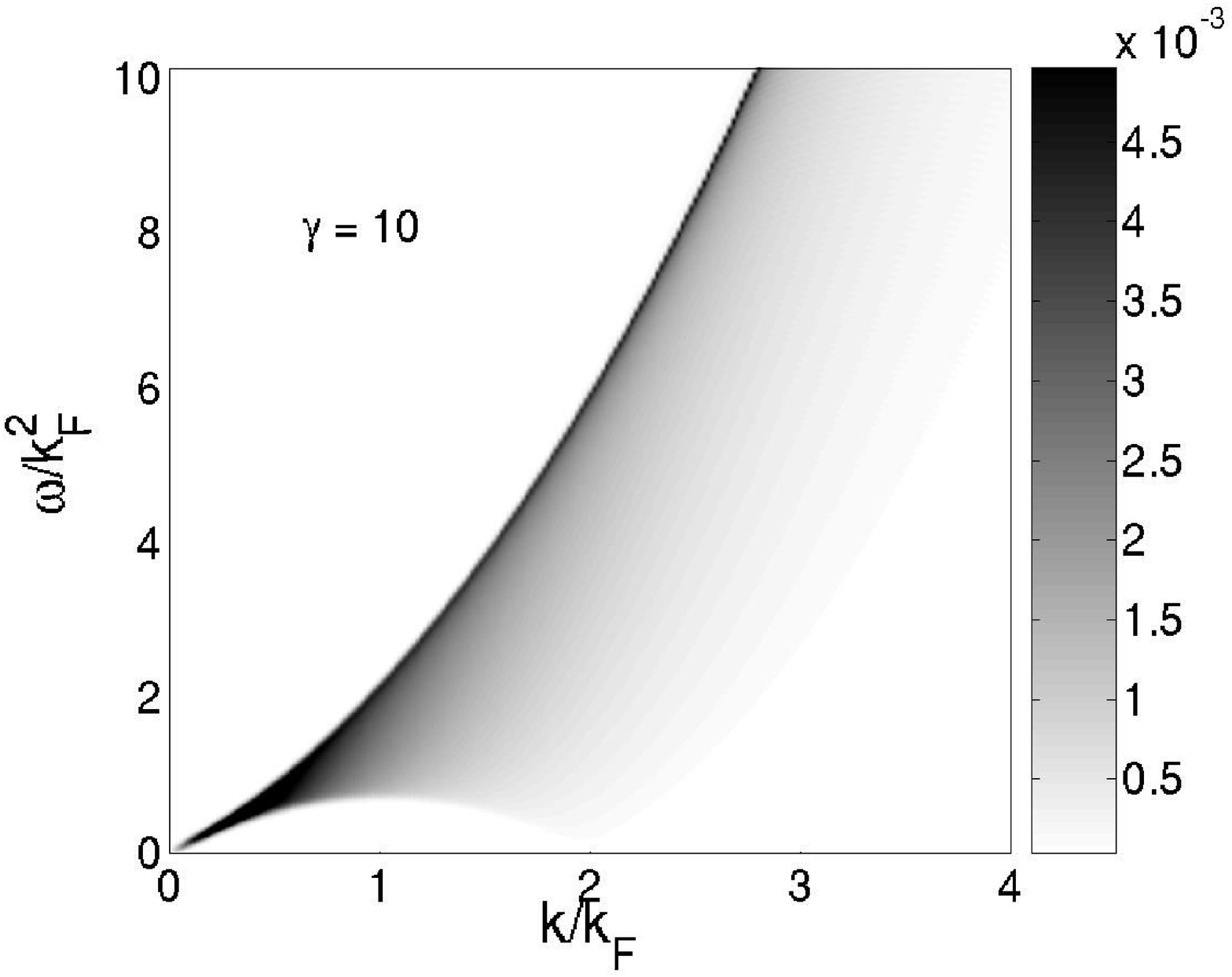} \\
\includegraphics[width=4.3cm]{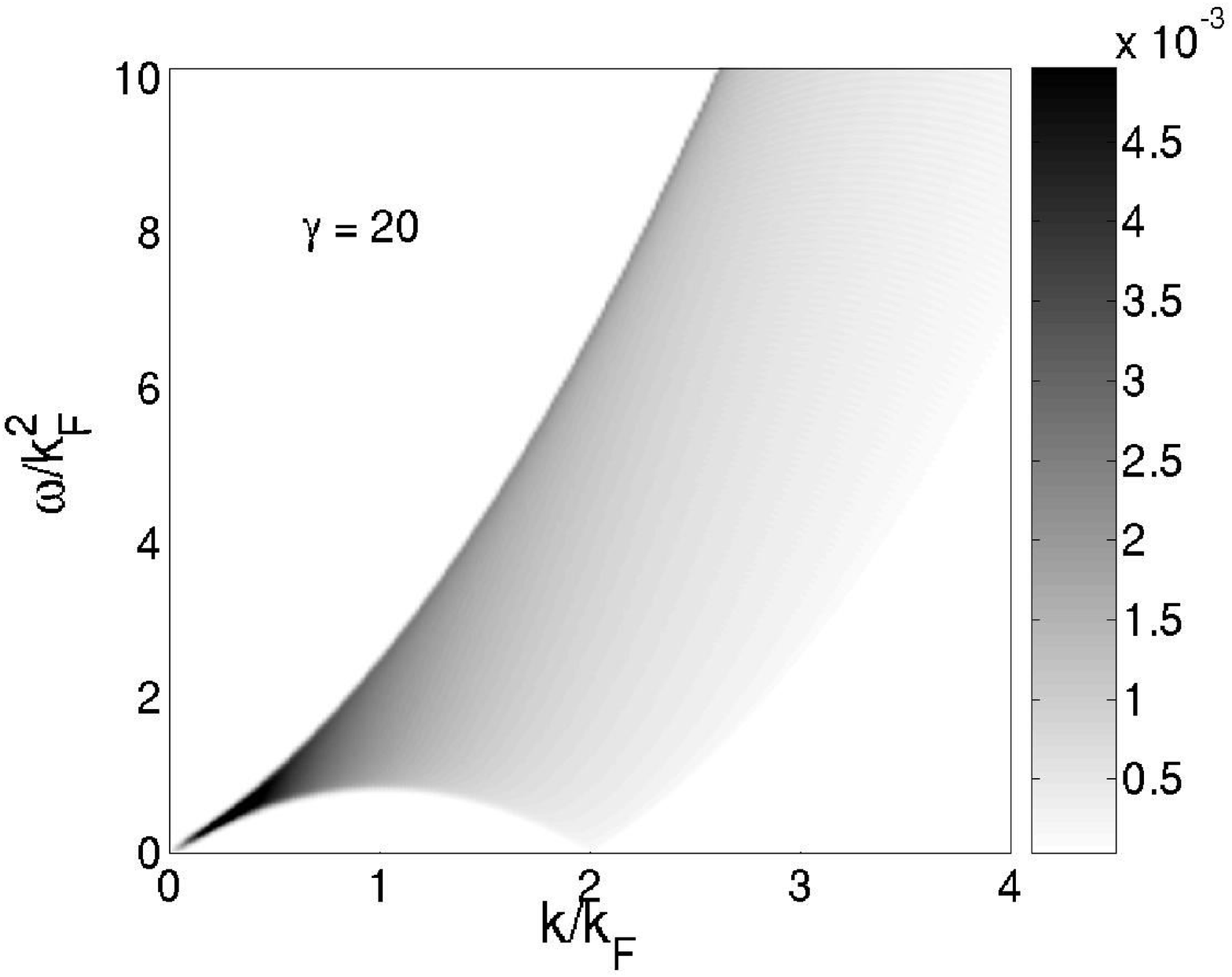}
&
\includegraphics[width=4.3cm]{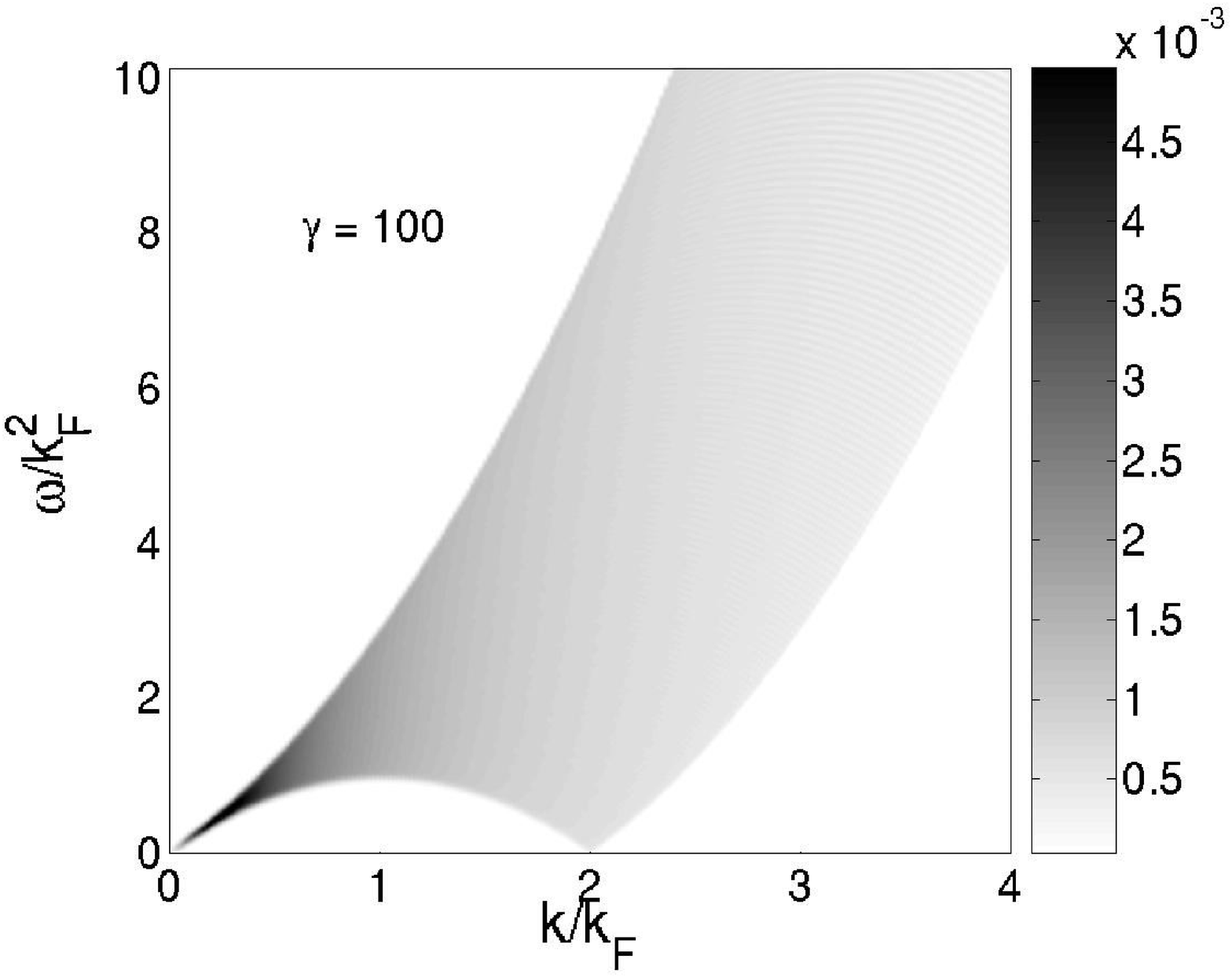}
\end{tabular}
\caption{Density plots of the dynamical structure factor as a function of interaction.
The horizontal axis is momentum
(running up to $4 k_F$), and the vertical axis represents energy transfer.  Data obtained from 
systems of length $L = 100$ with $N = 100$ and $\gamma = 0.25, 1, 5, 10, 20$ and $100$.}
\end{figure}

\begin{figure*}[!]
\begin{tabular}{ll}
\includegraphics[width=8cm]{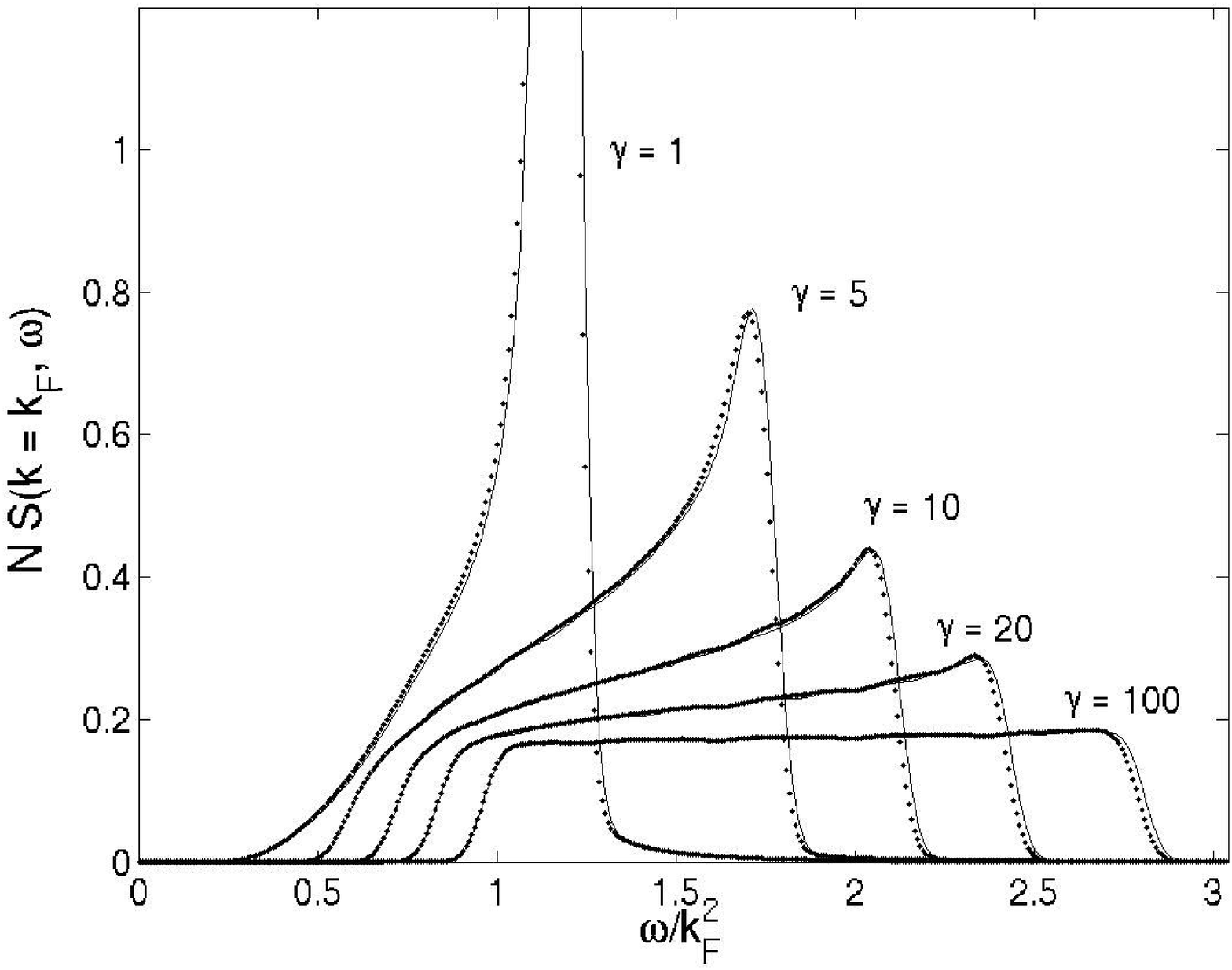} &
\includegraphics[width=8cm]{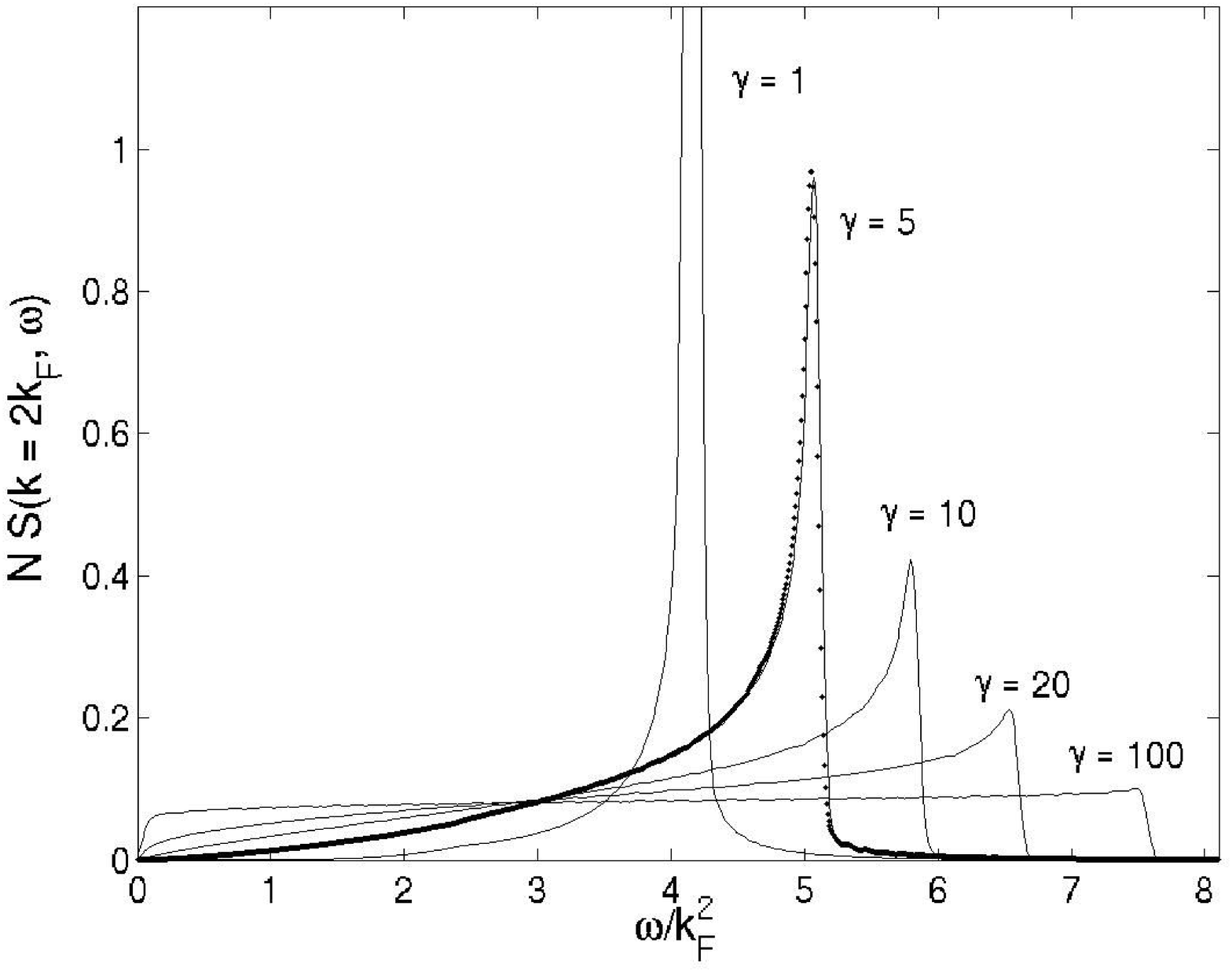} 
\end{tabular}
\caption{Fixed momentum plots of the dynamical structure factor for five representative
values of the interaction parameter $\gamma$ obtained using $L = 100$ and $N = 100$
(continuous curves) and $L = 80$ and $N = 80$ (dots), such that $k_F = \pi$ in all cases.  
The energy $\delta$-function in equation (\ref{SFsum}) is a Gaussian of width $w = 0.3$.}
\end{figure*}

By inserting a summation over intermediate states, $S (k, \omega)$ is transformed
into a sum of matrix elements of the density operator in the basis of Bethe
eigenstates $|\alpha\rangle$,
\begin{equation}
S (k, \omega) = \frac{2\pi}{L} \sum_{\alpha} |\langle 0 | \rho_k | \alpha \rangle|^2 
\delta (\omega - E_{\alpha} + E_0)
\label{SFsum}
\end{equation}
where $\rho_k = \sum_{j=1}^N e^{-ikx_j}$.  
All of the elements in each term of this sum are fully determined
by the Algebraic Bethe Ansatz, for a finite
system with specified boundary conditions (we consider periodic systems
in the present work).  
The Fock space is spanned by the set of Bethe wavefunctions, each 
fully determined by a set of $N$ rapidities $\{ \lambda_j \}$, solution
to the Bethe equations
\begin{equation}
\lambda_j + \frac{1}{L} \sum_k 2\arctan \frac{\lambda_j - \lambda_k}{c}
= \frac{2\pi}{L} I_j
\end{equation}
in which
$I_j$ are half-odd integers (integers) for even (odd) $N$.  
All solutions to these Bethe equations are real.  
Each set of 
distinct quantum numbers $\{ I_j \}$, $I_j \neq I_k$ if $j \neq k$ defines a Bethe
eigenstate participating in the sum (\ref{SFsum}).  The energy and 
momentum of such a state are given by $E = \sum_j \lambda_j^2$ and $k = \sum_j \lambda_j
= \frac{2\pi}{L} \sum_j I_j$.  The ground-state itself
is obtained from the set $\{ I_j^0 \}$, with 
$I_j^0 = \frac{N+1}{2} - j$, $j = 1,...,N$.  The wavefunction of an
eigenstate is given by the Bethe Ansatz, and its norm by the determinant
of the Gaudin matrix \cite{GaudinBOOK,KorepinCMP86}.

Matrix elements of the density operator in the basis of Bethe
eigenstates were calculated with the Algebraic Bethe Ansatz 
in [\onlinecite{SlavnovTMP79_82}].  They are
given by the determinant of a matrix whose entries are rational
functions of the rapidities of the two eigenstates involved.  For the
sake of brevity we do not reproduce these expressions here.  

What remains to be performed is the actual summation over
intermediate states in (\ref{SFsum}).
From this step onwards, everything is done numerically.  The Fock space
of intermediate states is scanned by navigating through 
choices of sets of quantum numbers.  
For each individual intermediate state, the Bethe equations are solved,
and the matrix element is computed.  To obtain smooth curves in energy, 
the energy delta function in (\ref{SFsum}) is broadened to
a width equal to a multiple of the typical energy level spacing.  The 
contribution to the dynamical structure factor sum is tallied until
good convergence has been achieved.  This is quantified by evaluating
the $f$-sum rule,
\begin{equation}
\int \frac{d\omega}{2\pi} \omega S(k, \omega) = \frac{N}{L} k^2.
\label{fsumrule}
\end{equation}
Since this is skewed towards high energy, and in view of the ordering 
of states in the scanning we perform (typically going from low-energy intermediate
states to higher-energy ones), the saturation level of this 
sum rule represents a lower bound for the saturation of $S(k, \omega)$ itself.

\begin{table}
\caption{$f$-sum rule saturation percentage achieved at the two 
representative values of momentum used in Fig. 2.  The method
converges fastest in the strongly-interacting regime, the most difficult
regime being intermediate interactions.}
\begin{tabular}{|c|c|c|c|c|c|c|}
\hline
$\gamma$ & 0.25 & 1 & 5 & 10 & 20 & 100 \\ \hline \hline
$k_F$ & 99.52 & 99.44 & 99.48 & 99.45 & 99.80 & 99.97 \\ \hline
$2k_F$ & 99.34 & 99.11 & 97.49 & 98.21 & 99.35 & 99.90 \\ \hline
\end{tabular}
\label{SRtable}
\end{table}

\begin{table}
\caption{Sound velocities for the interaction values 
considered here.  A comparison is made between the finite-size
value, and the infinite-size one.}
\begin{tabular}{|c||c|c|c|c|c|c|}
\hline
$\gamma$ & 0.25 & 1 & 5 & 10 & 20 & 100 \\ \hline \hline
$v_s (N = 100)$ & 0.9612 & 1.8350 & 3.5916 & 4.4818 & 5.2175 & 6.0395 \\ \hline
$v_s (\infty)$ & 0.9594 & 1.8342 & 3.5912 & 4.4816 & 5.2173 & 6.0395 \\ \hline 
\end{tabular}
\label{vstable}
\end{table}

It is useful
to recall here the nature of excitations in the Lieb-Liniger model, which come in
two types \cite{LiebPR130},
Type I (``particles'') and Type II (``holes'') \cite{Exc}.  
Type I
are Bogoliubov-like quasiparticles that 
exist for any momentum, and represent states with one quantum
number displaced outside the ground-state interval.  Their 
dispersion relation is described in the thermodynamic limit by 
an integral equation, yielding a curve contained
between the asymptotic limits $\epsilon_I (k) = k^2$ at $\gamma = 0$
and $\epsilon_I (k) = k^2 + 2\pi n |k|$ for $\gamma \rightarrow \infty$.
Type II excitations are holes in the ground-state distribution, and
do not appear in Bogoliubov theory.  
They exist in the interval $|k| \leq k_F \equiv \pi n$, and 
their dispersion relation coincides with 
the lower threshold of the DSF.
Low-energy Umklapp modes at $k = 2k_F$ can be understood as 
excitations going from one side of the Fermi surface 
to the other.  
Both Type I and Type II dispersion relations approach $k \rightarrow 0$
with a slope equal to the velocity of sound $v_s (\gamma)$.  

\begin{figure}
\includegraphics[width=8cm]{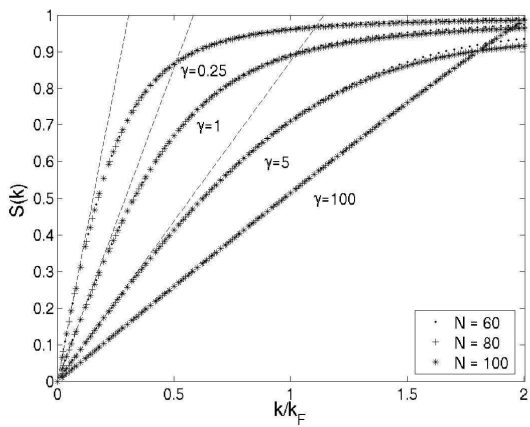} 
\caption{Static structure factor for four representative
values of the interaction parameter $\gamma$.  The three
values of system size illustrate the smallness of finite-size
effects.    
All curves saturate towards the limit $S(k \rightarrow \infty) = 1$.
The dashed lines are the small momentum asymptotic predictions
$S(k) \simeq k/v_s$.  Deviations in the $\gamma = 5$ curve near
$2k_F$ are commensurate with lack of saturation of the f-sum
rule there (see Table I).}
\end{figure}

A remarkable feature in this method comes from the fact that
it is not necessary to scan through the whole Fock space to get good
saturation.  Contributions from intermediate states with up to only a
handful of particles are sufficient to achieve extremely good accuracy.
The saturation of the $f$-sum rule (\ref{fsumrule}) achieved for the $N = 100$ curves
in Fig. 2 is summarized in Table 1 (for lower values of $N$, even better
saturation is achieved).  Our algorithm is designed to
recursively hunt for the most important terms in the multiparticle sum, 
in decreasing order of contribution to the DSF, and therefore to
maximize the efficiency of the available numerical resources.
Higher accuracy is obtained by using more computational time to
include the contributions from more intermediate states.

Fig. 1 shows density plots for the DSF obtained
with our method, for different values of $\gamma$.
We used unit density $n = 1$ with $L = 100$ and $N = 100$ (this
compares with experimental particle numbers \cite{KinoshitaSCIENCE305,ParedesNATURE429}).
For small $\gamma$, the DSF is essentially a single delta peak centered on the
Type I dispersion relation,
$S (k, \omega) = \frac{N k^2}{L \epsilon_I(k)} \delta (\omega - \epsilon_I(k))$.
As $\gamma$ increases, the peak flattens but remains near the upper
two-particle boundary.
In particular, this confirms that the peak near the lower boundary observed 
in first-order RPA is an artefact of that method \cite{BrandPRA72_73}.  
Increasing $\gamma$ further towards the Tonks-Girardeau limit, the DSF approaches a constant over a finite 
frequency interval for a given momentum.  All of this is illustrated more specifically
for two specific values of momentum in Fig. 2.  Fig. 3 shows $S(k)$, which qualitatively fits with quantum
Monte Carlo results \cite{AstrakharchikPRA68}.  At small momentum,
we recover the prediction 
$S(k) \simeq k/v_s$
(see for example [\onlinecite{AstrakharchikPRA70}]).
The curve for the largest interaction value, $\gamma = 100$, also clearly approaches the well-known
result $\lim_{\gamma \rightarrow \infty} S(k) = k/2k_F$ for $k \leq 2k_F$.
Moreover, Fig. 1 shows that low-energy contributions near $2k_F$, which represent 
superfluidity-breaking Umklapp modes, are only important for large $\gamma$.

For all values of $\gamma$, the signal mostly lies between the 
two-particle continuum defined by convolution of the Type I and Type II dispersion
relations.  For general $k$, all the signal in fact lies 
strictly above the Type II dispersion relation modulo $2k_F$ translations, 
since there are no lower-energy states
available.  The upper two-particle bound is however not robust: 
multiparticle contributions give a nonzero signal at (in principle arbitrarily)
large energy (coming from {\it e.g.} states with two or more particles of large
but opposite momentum).  In practice, however, the data shows that the onset
of the DSF at finite $\gamma$ is followed by a sharp peak around the 
Type I dispersion, followed by a rapid decrease.  We believe that in 
the thermodynamic limit, contributions from intermediate states with higher particle numbers smoothen
the upper threshold into a high-frequency tail, as is the case for the
corresponding correlators in quantum spin chains
\cite{Pereira0603681}.  The only exception
is the Tonks-Girardeau limit, where both the lower and
upper thresholds remain sharp.

To quantify finite-size effects in our results, we have included data for
$N = 80$ in Fig. 2 and for $N = 60, 80$ in Fig. 3
and given values for the effective $v_s (\gamma)$ (the 
conformal exponent $\eta$, such that $S (2k_F, \omega) \simeq \omega^{\eta - 2}$ 
at small $\omega$, is given by $\eta = 4\pi n /v_s$) obtained for $N = 100$ compared to the
thermodynamic one in Table \ref{vstable}.  Theoretical considerations based on
Bethe Ansatz expressions for correlation functions \cite{KitanineNPB554}, similar
to those used here, predict finite-size corrections of order $1/N$.  
The precise form of these $1/N$ corrections then depends on the specific boundary
conditions used, but we believe that our results are close enough to the thermodynamic limit
to make the choice of boundary conditions immaterial.  Specifically, in Fig. 2,
the only observable effect of increasing system size is a very slight shift of the main peaks of the 
DSF towards higher energy (for clarity we do not plot the $N = 80$ results in the right-hand figure, but
they show the same behaviour as in the left-hand one).  
The static structure factor plotted in Fig. 3
shows essentially no change for the different values of $N$ given.  The variations
observed fit comfortably within the deviation from perfection of the sum rule saturation achieved,
which is the actual determining factor in the quality of our results.  More important 
for theory is the question of a confining potential 
\cite{AstrakharchikPRA68,BatrouniPRA72,GattobigioJPB39}, which is present in experiments but 
breaks the integrability of the Lieb-Liniger model.  We expect
that experiments on large enough systems would on the other hand show correlations 
approaching those of a pure Lieb-Liniger model similar to those obtained here, or
that experiments could be done in box-like geometries, where a variation of our method
would be applicable.

Summarizing, we have computed the frequency- and momentum-dependent 
dynamical density-density correlation
function of the one-dimensional interacting Bose gas (Lieb-Liniger model)
for systems with finite numbers of particles, using a Bethe Ansatz-based
numerical method.  This goes beyond other available methods in offering
a full characterization of the momentum and frequency dependence of the
dynamical structure factor, provides a firm testing standard for other methods, and 
opens the way to many possible extensions (other correlators,
systems with mixed statistics, finite temperatures) on which we will report
in future publications.

J.-S. C. acknowledges useful discussions with N. A. Slavnov, M. J. Bhaseen,
G. V. Shlyapnikov 
and J. T. M. Walraven.  P. C. acknowledges discussions with M. Polini.
This research was supported by the Stichting voor Fundamenteel Onderzoek
der Materie (FOM) of the Netherlands.


\begin{thebibliography}{99}
\bibitem{PitaevskiiBOOK} L. Pitaevskii and S. Stringari, ``Bose-Einstein Condensation'',
Oxford, 2003.
\bibitem{PetrovPRL85} D. S. Petrov, G. V. Shlyapnikov and J. T. M. Walraven, Phys. Rev. Lett.
{\bf 85}, 3745 (2000).
\bibitem{GoerlitzPRL87} A. G\"orlitz {\it et al}., Phys. Rev. Lett. {\bf 87}, 130402 (2001).
\bibitem{GreinerPRL87} M. Greiner {\it et al}., Phys. Rev. Lett. {\bf 87}, 160405 (2001).
\bibitem{MoritzPRL91} H. Moritz {\it et al}., Phys. Rev. Lett. {\bf 91}, 250402 (2003).
\bibitem{ParedesNATURE429} B. Paredes {\it et al}., Nature {\bf 429}, 277 (2004).
\bibitem{KinoshitaSCIENCE305} T. Kinoshita, T. Wenger and D. S. Weiss, Science {\bf 305}, 1125 (2004).
\bibitem{GangardtPRL90} D. M. Gangardt and G. V. Shlyapnikov, Phys. Rev. Lett. {\bf 90}, 010401 (2003);
K. V. Kheruntsyan
{\it et al}., 
Phys. Rev. A {\bf 71}, 053615 (2005).
\bibitem{LaburthePRL92} B. Laburthe Tolra {\it et al}., Phys. Rev. Lett. {\bf 92}, 190401 (2004).
\bibitem{LiebPR130} E. H. Lieb and W. Liniger, Phys. Rev. {\bf 130}, 1605 (1963);  
E. H. Lieb, Phys. Rev. {\bf 130}, 1616 (1963).
\bibitem{TonksPR50} L. Tonks, Phys. Rev. {\bf 50}, 955 (1936).
\bibitem{GirardeauJMP1} M. Girardeau, J. Math. Phys. (N.Y.) {\bf 1}, 516 (1960).
\bibitem{YangJMP10} C. N. Yang and C. P. Yang, J. Math. Phys. (N.Y.) {\bf 10}, 1115 (1969).
\bibitem{ZambelliPRA61} F. Zambelli {\it et al}., Phys. Rev. A {\bf 61}, 063608 (2000).
\bibitem{HaldanePRL47} F. D. M. Haldane, Phys. Rev. Lett. {\bf 47}, 1840 (1981).
\bibitem{BerkovichPLA142} A. Berkovich and G. Murthy, Phys. Lett. A {\bf 142}, 121 (1989).
\bibitem{CastroNetoPRB50} A. H. Castro Neto et al., Phys. Rev. B {\bf 50}, 14032 (1994).
\bibitem{LuxatPRA67} D. L. Luxat and A. Griffin, Phys. Rev. A {\bf 67}, 043603 (2003).
\bibitem{OlshaniiPRL91} M. Olshanii and V. Dunjko, Phys. Rev. Lett. {\bf 91}, 090401 (2003).
\bibitem{BrandPRA72_73} J. Brand and A. Yu. Cherny, Phys. Rev. A {\bf 72}, 033619 (2005);
A. Yu. Cherny and J. Brand, Phys. Rev. A {\bf 73}, 023612 (2006).
\bibitem{PolletPRL93} L. Pollet et al, Phys. Rev. Lett. {\bf 93}, 210401 (2004).
\bibitem{AstrakharchikPRA68} G. E. Astrakharchik and S. Giorgini, Phys. Rev. A {\bf 68}, 031602(R) (2003).
\bibitem{KorepinBOOK} V. E. Korepin, N. M. Bogoliubov and A. G. Izergin, ``Quantum Inverse Scattering
Method and Correlation Functions'', Cambridge, 1993, and references therein.
\bibitem{KorepinCMP94} V. E. Korepin, Commun. Math. Phys. {\bf 94}, 93 (1984).
\bibitem{SlavnovTMP79_82} N. A. Slavnov, Teor. Mat. Fiz. {\bf 79}, 232 (1989); {\it ibid}., {\bf 82}, 389 (1990).
\bibitem{CauxPRL95} J.-S. Caux and J. M. Maillet, Phys. Rev. Lett. {\bf 95}, 077201 (2005);
J.-S. Caux, R. Hagemans and J. M. Maillet, J. Stat. Mech. (2005) P09003.
\bibitem{DuanPRL96} L.-M. Duan, Phys. Rev. Lett. {\bf 96}, 103201 (2006).
\bibitem{StengerPRL82} J. Stenger {\it et al}., Phys. Rev. Lett. {\bf 82}, 4569 (1999).
\bibitem{KorepinCMP86} V. E. Korepin, Commun. Math. Phys. {\bf 86}, 391 (1982).
\bibitem{GaudinBOOK} M. Gaudin, ``La fonction d'onde de Bethe'', Masson (Paris) (1983).
\bibitem{Exc} Although useful for visualizing low-energy excitations,
this classification yields double counting (Type II particles are really
a compound of low-momentum Type I particles \cite{LiebPR130}).  
The Fock space itself is spanned by using only Type I.
\bibitem{AstrakharchikPRA70} G. E. Astrakharchik and L. Pitaevskii,
Phys. Rev. A {\bf 70}, 013608 (2004).
\bibitem{KitanineNPB554} N. Kitanine, J. M. Maillet and V. Terras, Nucl. Phys. {\bf B554}, 647 (1999).
\bibitem{Pereira0603681} R. G. Pereira {\it et al}., cond-mat/0603681.
\bibitem{BatrouniPRA72} G. G. Batrouni et al,  Phys. Rev. A {\bf 72}, 031602 (2005).
\bibitem{GattobigioJPB39} M. Gattobigio, J. Phys. B {\bf 39}, S191 (2006).
\end{thebibliography}
\end{document}